\documentclass[pre, twocolumn,showpacs,preprintnumbers,amsmath,amssymb ]{revtex4}

\usepackage{graphicx}
\usepackage{dcolumn}
\usepackage{bm}
\usepackage{float}
\usepackage{amsmath}
\usepackage{subfigure}
\usepackage[percent]{overpic}
\usepackage{color}

\begin{document}

\input epsf.sty

\title{Dimensionality of spin modulations in 1/8-doped lanthanum cuprates from the perspective of NQR and $\mu$SR experiments}
\author{J. Gerit Brandenburg
   and  Boris V. Fine
  }
\affiliation{Institute for Theoretical Physics, University of Heidelberg, Philosophenweg 19, 69120 Heidelberg, Germany}

\date{\today}

\begin{abstract}
We investigate the dimensionality of inhomogeneous spin modulation patterns in the cuprate family of high-temperature superconductors with particular focus on 1/8-doped lanthanum cuprates. We compare one-dimensional stripe modulation pattern with two-dimensional chackerboard of spin vortices in the context of nuclear quadrupole resonance(NQR) and muon spin rotation($\mu$SR) experiments. In addition, we also consider the third pattern, a two-dimensional superposition of spin spirals.
Overall, we have found that none of the above patterns leads to a consistent interpretation
of the two types of experiments considered. This, in particular, implies that the
spin vortex checkerboard cannot be ruled out on the basis of available NQR/$\mu$SR experimental results.
\end{abstract}

\maketitle

\section{Introduction}
\label{Sec:Introduction}

The apparent connection between superconductivity and magnetism in cuprates, pnictides and other unconventional superconductors is not, at present, well understood theoretically. A possible mechanism behind this connection is through the formation of slowly fluctuating inhomogeneous magnetic structures emerging as a result of strong tendency towards electronic phase separation\cite{Fine2008}. Electronic phase separation fluctuating in space and time is difficult to investigate experimentally. In this respect, the lanthanum family of high temperature cuprate superconductors at the doping level 1/8 has long been the subject of intense interest, due to the fact that electronic spin and charge modulations in these materials are static in time and periodic in space.  The character of these modulations, however, is still the subject to mutually exclusive propositions. The purpose of the present paper is to examine these propositions from the viewpoint of nuclear quadrupole resonance (NQR) experiments\cite{Hunt2001} and muon spin rotation ($\mu$SR) experiments\cite{Luke1997,Nachumi1998,Kojima2000}. 

At present, the dominant viewpoint is that the spin and charge modulations in 1/8-doped lanthanum cuprates have the character of one-dimensional stripes\cite{Tranquada1995,Tranquada1996,Kivelson2003}. Motivated by the STM observations of checkerboards\cite{Hoffman2002,Hanaguri2004,Vershinin2004,Levy2005}, one of us previously investigated the two-dimensional alternatives to the stripe interpretations --- initially in the form of grid\cite{Fine2004,Fine2005,Fine2007NQR} and then, after neutron scattering evidence against grid appeared\cite{Christensen2007}, in the form of spin vortex checkerboard\cite{Fine2007,Fine2011}. Refs.\cite{Fine2004,Fine2007NQR,Fine2007,Fine2011} also contain detailed discussion of the arguments in favor and against the stripe proposition. The readers are referred to Refs.\cite[]{Seibold1998,Berciu1999,Timm2000,Mitsen2004,Wilson2006,Wilson2008,Wilson2009,Koizumi2008,Azzouz2010} for various considerations pertaining to the formation of spin vortices and other non-collinear spin  superstructures.

Static spin modulations create the distribution of local magnetic fields, whose effect is observable by both NQR and $\mu$SR experiments. The advantage of these experiments is that one can make experimental predictions for a given spin modulation pattern essentially without further theoretical assumptions. The disadvantage is that these techniques do not  resolve the wave vectors of spin modulations. 

The comparative analysis of the stripe and grid superstructures from the viewpoint of NQR experimentswas done in Refs.\cite{Fine2007NQR} and from the viewpoint of $\mu$SR in Ref.\cite{Kojima2000}. The results of the NQR analysis indicated that the stripe pattern would lead to significant discrepancies with experiment due to a certain singularity in the distribution of local magnetic fields. At the same time, the grid pattern  did not lead to that singularity and, as a result, was better compatible with NQR experiments. In general, however, the NQR experiments were even better compaible with a totally random Gaussian distribution of the local fields, which, in turn, would be incompatible with the static modulated spin response observed by elastic neutron scattering\cite{Tranquada1995,Tranquada1996}.
The $\mu$SR results\cite{Kojima2000}, on the other hand, were more consistent with the stripe pattern than with the grid precisely because of the presence of the above mentioned singularity on the distribution of local magnetic fields. At the same time, it was not possible to reproduce the $\mu$SR results quantitatively from the full three-dimensional stacking of either stripes or grid.

The above rather incoherent phenomenology is further complicated by the discrepancy in the estimate of the spin modulation amplitude from $\mu$SR (about $0.3 \mu_B$) and from neutron scattering ($0.1-0.15 \mu_B$). The NQR experiments are more consistent with the latter estimate.

So far, the attempts to analyze NQR/$\mu$SR data have not included the spin vortex lattice. It is our goal in this paper to check whether this pattern may help to reconcile NQR/$\mu$SR/neutron scattering phenomenology. For completeness of the analysis, we have also done NQR/$\mu$SR calculations for a coherent superposition of two orthogonal spiral harmonics previously mentioned in Ref.\cite{Fine2007}. We further investigated possible muon sites which are different from the ``standard'' site identified in Ref.\cite{Hitti1990}.

\begin{figure*}[t]
 \centering

      \hfill
      \begin{overpic}[width=4.5cm
]{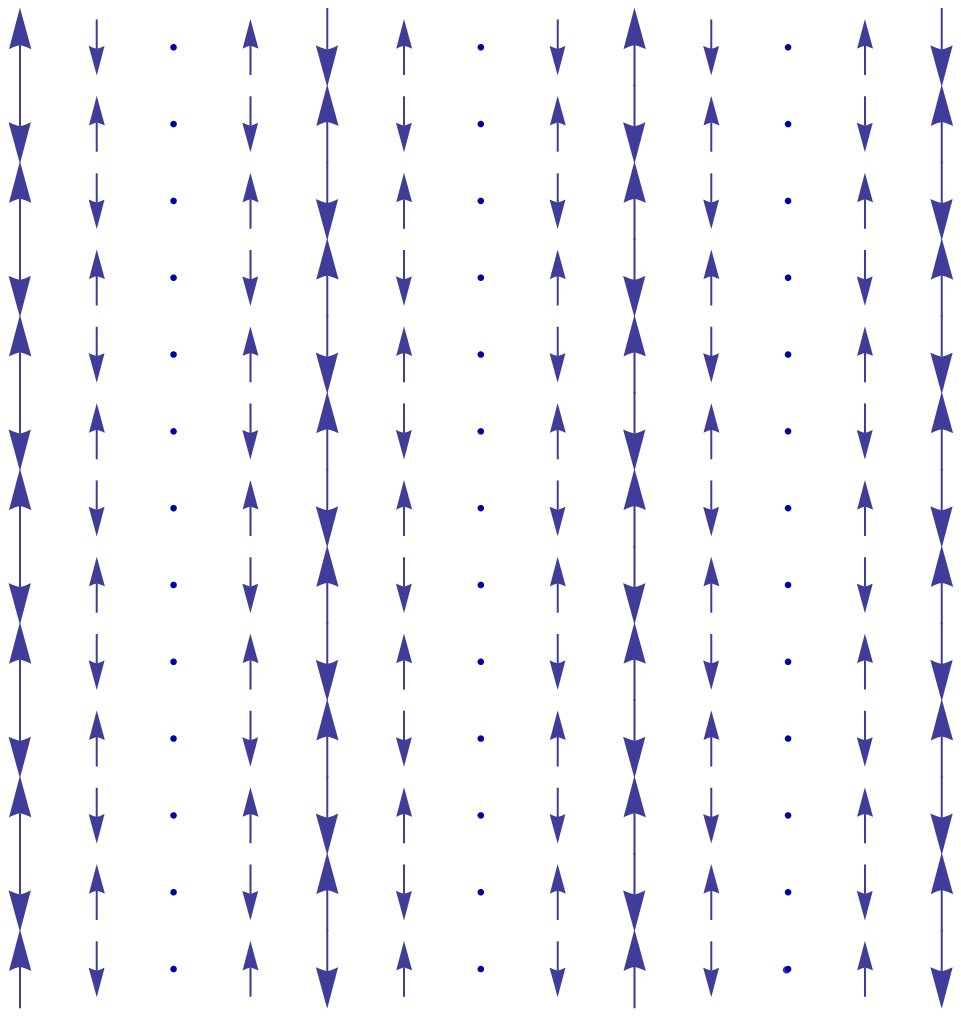}
	\put(-10,95){\textcolor{black}{(a)}}
      \end{overpic}
      \hfill
      \begin{overpic}[width=4.5cm]{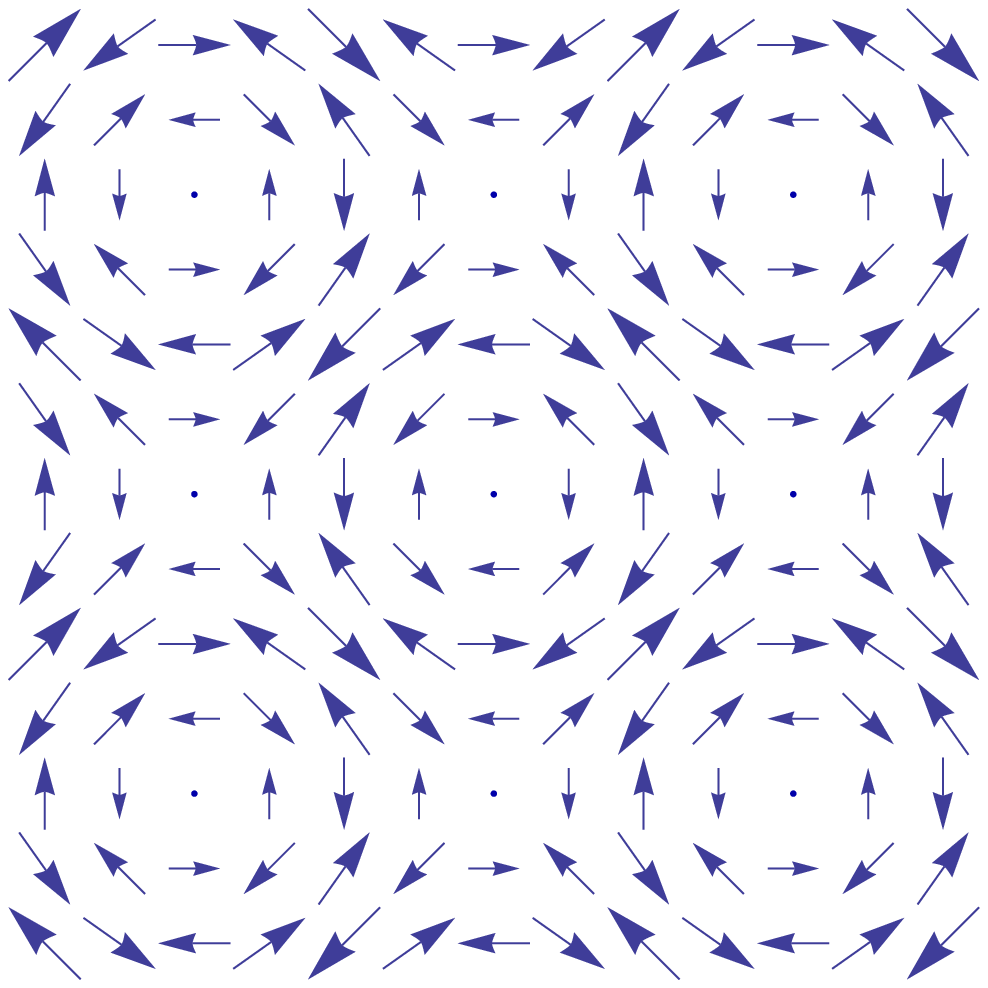}
	\put(-10,95){\textcolor{black}{(b)}}
      \end{overpic}
      \hfill
      \begin{overpic}[width=4.5cm]{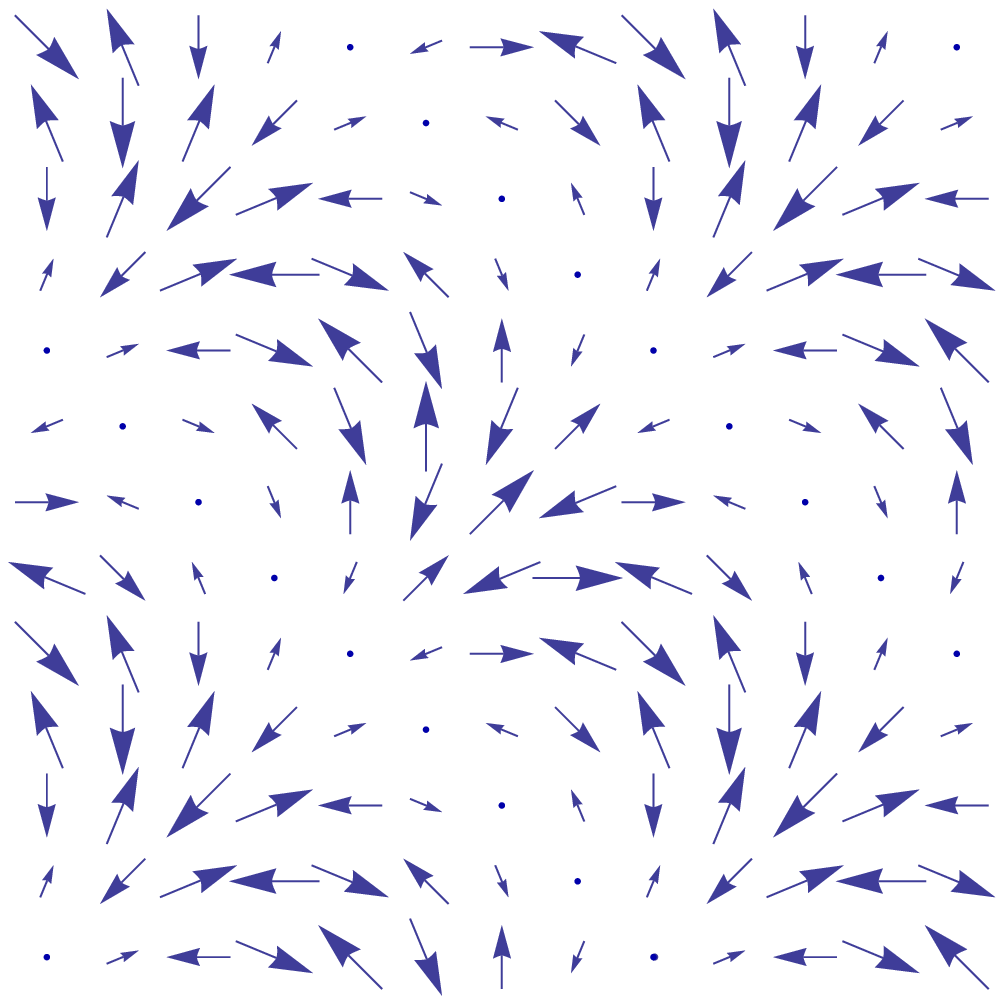}
	\put(-10,95){\textcolor{black}{(c)}}
      \end{overpic}
      \hfill

    \caption{Site-centered, commensurate spin superstructures (a) stripes, (b) spin-vortex lattice, and (c) 2D superposition of spin spirals.}
    \label{Spinmodulation}
\end{figure*}

\section{Theoretical Formulation}
\label{Sec:TheoreticalFormulation}
\subsection{Spin Structures}
\label{Subsec:SpinStructures}
In 1/8-doped lanthanum cuprates, the primary experimental evidence\cite{Tranquada1995} of spin modulations is based on the four-fold splitting of the antiferromagnetic $\left({\pi \over a} \pm \delta {2\pi \over a},{\pi \over a} \pm \delta {2\pi \over a}\right)$, were $a$ the in-plain lattice period and $\delta \approx 1/8$. This splitting indicates that the antiferromagnetic order is modulated with period $a/\delta$. The experiments also indicate that spin polarizations mainly lie ``in-plain''. All possible modulations that may be consistent with the above four peaks can be represented as
\begin{eqnarray}
\mathbf S_{ij} = (-1)^{i+j} &
\left[
\; \;
\left(
\begin{array}{c}
S_{1x} \hbox{cos} \left( \mathbf q_1 \cdot \mathbf r_{ij} + \varphi_{1x} \right) \\
S_{1y} \hbox{cos} \left( \mathbf q_1 \cdot \mathbf r_{ij} + \varphi_{1y} \right)
\end{array}
\right)
 \right.
\nonumber
\\
&
\; \;
\left.
+
\left(
\begin{array}{c}
S_{2x} \hbox{cos} \left( \mathbf q_2 \cdot \mathbf r_{ij} + \varphi_{2x} \right) \\
S_{2y} \hbox{cos} \left( \mathbf q_2 \cdot \mathbf r_{ij} + \varphi_{2y} \right)
\end{array}
\right)
\right],
\label{S}
\end{eqnarray}
where $i$ and $j$ are the indices of the square lattice, $r_{ij}$ are the positions of the lattice sites, $\mathbf S_{ij}$ the static spin polarization values for each site, $\mathbf q_1 = \left( \delta {2\pi \over a},0 \right)$ and $\mathbf q_2 = \left(0, \delta {2\pi \over a} \right)$ are the modulation wave vectors,  $\{ S_{1x},S_{1y},S_{2x}S_{2y}\}$ and 
$\{\varphi_{1x},\varphi_{1y},\varphi_{2x},\varphi_{2y}\}$ are respectively the amplitudes and the phases of four possible linear polarization harmonics. Below, we use variable $S_0$ to denote the maximum spin polarization of a given pattern. We also use Bohr magneton units for electronic spin polarizations, so that the maximum polarization of spin 1/2 corresponds to $S_0 = \mu_B$.

Possible specific patterns of general form(\ref{S})include: (i) stripes, e.g.  $\{ S_{1x} = 0,S_{1y} = S_0, S_{2x}=0, S_{2y}= 0 \}$ [Fig.~\ref{Spinmodulation}(a)]; (ii) spirals, e.g.  $\{ S_{1x} = S_0,S_{1y} = S_0, S_{2x}=0, S_{2y}= 0 , \varphi_{1x}-\varphi_{1y}= \pi/2 \}$; (iii) grid, e.g. $\{ S_{1x} = 0,S_{1y} = S_0/2, S_{2x}=0, S_{2y}= S_0/2 \}$; (iv) spin vortex checkerboard, e.g. $\{ S_{1x} = 0,S_{1y} = S_0/\sqrt{2}, S_{2x}=S_0/\sqrt{2}, S_{2y}= 0 \}$ [Fig.~\ref{Spinmodulation}(b)]; (v) Superposition of two orthogonal spirals, e.g. $\{ S_{1x} = S_0/2,S_{1y} = S_0/2, S_{2x}=S_0/\sqrt{2}, S_{2y}= S_0/\sqrt{2}, \varphi_{1x}-\varphi_{1y}= \pi/2, \varphi_{2x}-\varphi_{2y}= \pi/2 \}$ [Fig.~\ref{Spinmodulation}(c)]. All these patterns are locally stable in the approximation of staggered spin polarizations\cite{Fine2007}. The stripe and the spiral interpretations are one-dimesional and, therefore, require two kinds of domains with orthogonal modulations to account for the four-fold splitting of the antiferromagnetic peak.

Neutron scattering study of Ref.~\cite{Christensen2007} indicates that the above spin modulation harmonics should be transversely polarized, which implies that $ S_{1x} = 0$ and $ S_{2y} = 0$. This, in turn, is inconsistent with the spiral patterns (iii) and (v) and with the grid pattern (iii), thus leaving the transversely polarized stripes (i) and spin vortex superlattice (iv) as the only two remaining propositions. The comparative analysis of these two propositions is the primary goal of this work. Below, however, we also present NQR/$\mu$SR calculations for the two-spiral pattern (v), which were done, in part, to make the analysis complete and, in another part, because as obvious from Fig.~\ref{Spinmodulation}(c)] this pattern appears to be consistent with the minimum of the pseudogap along the diagonal lattice direction\cite{Valla2006}.

When necessary for the $\mu$SR calculations, the three-dimensional arrangement of the above spin modulations was chosen as follows. For stripes, we use the orthogonal stripe directions in the adjacent layers, and in second adjacent layer --- half-a-period parallel shift. Spin vortices checkerboards and the two-spiral superstructures are both shifted in the adjacent layers by approximately two lattice periods in both the $a-$ and $b$-directions, so that the spin-poor regions in one layer are located on the top of the spin-rich regions in the adjacent layer.  In the case of stripes, the above arrangement is what is proposed by Tranquada, even though theoretical doubts about such proposition exist\cite{Fine2004,Fine2011}. For the two other superstractures, the above arrangement is based solely on the basis of Coulomb repulsion between the same-charge regions, and the assumption that spin-poor regions accumulate charge carriers and thus positively charged, while spin-rich regions are negatively charged.

In the following, we also explore the noisy modifications of the above modulations. The spin noise is to be uncorrelated between different lattice sites. For each site, it will have two in-plane components, each randomly selected from the Gaussian distribution with zero average and root-mean-squared deviation denoted as $\Delta S$. The noise was incorporated in the NQR calculations by analytically modifying the formulas presented in the next section, while the $\mu$SR calculations were based on the actual sampling of the three dimensional spin structure with or without the noise.

\subsection{NQR lineshapes}
\label{Subsec:NQRlineshapes}

At low temperatures, static spin modulations lead to a broad distribution of local hyperfine fields that affect NQR frequencies and transition rates. In this paper, we attempt to describe the experimental low-temperature $^{63}$Cu NQR lineshape reported in Ref.\cite{Hunt2001}. 
Nuclear isotopes $^{63}$Cu have spin $3/2$. The NQR Hamiltonian for each nuclear spin is assumed to have form
\begin{equation}
{\cal H} = \frac{\nu_Q h}{2} \left\{I_z^2 - \frac{1}{3} I (I+1)\right\}- \gamma_{\textrm{Cu}} h {\mathbf H \cdot \mathbf I}
\label{Hamiltonian}
\end{equation}
where $\mathbf I$ is the nuclear spin operator, $h$ is the Planck's constant, $\gamma_{\textrm{Cu}}$ is the gyromagnetic ratio, $\mathbf H$ is the local hyperfine field created by static spin modulations,  $\nu_Q$ is the quadrupolar parameter, and the $z$-axis is perpendicular to the CuO$_2$ plane. The hyperfine local field at the positon $\mathbf r_{ij}$ is given by
\begin{equation}
\mathbf H (\mathbf r_{ij}) = A \ \mathbf S_{ij} + B \ \sum^{\hbox{\tiny NN}}_{kl} \mathbf S_{kl},
\label{Hhyper}
\end{equation}
where $A=38$ [kOe/$\mu_B$] and $B=42$ [kOe/$\mu_B$], and the sum is taken over the four nearest neighbors of spin $\mathbf S_{ij}$.

The calculation of the NQR lineshape (echo intensity as a function of the frequency of NQR echo pulses) requires computing six transition frequencies between the four eigenstates of the Hamiltonian (\ref{Hamiltonian}), weighing these transitions by the square of the appropriate matrix element and by the square of the frequency itself and then averaging over all possible values of the hyperfine field associated with a given spin superstructure. 
The framework and the NQR parameters of our calculations including the A- and B- NQR lines and broadening of these lines are identical to those used in Refs.\cite{Hunt2001,Fine2007NQR}, where stripe and grid superstructures were considered.
Here we only describe the last step associated with computing the hyperfine field distribution due to the spin vortex checkerboard [Fig.~\ref{Spinmodulation}(b)] and the two-spiral modulation[Fig.~\ref{Spinmodulation}(c)]. We assume that the periods of both modulations are very close but not exactly equal to 8 lattice periods. As a result, both modulations are incommensurate, and the resulting distribution of hyperfine fields involves the averaging over the ``ergodic'' phase of these modulations.

\begin{figure*}[t]
 \centering
      \hfill
      \begin{overpic}[height=4cm
]{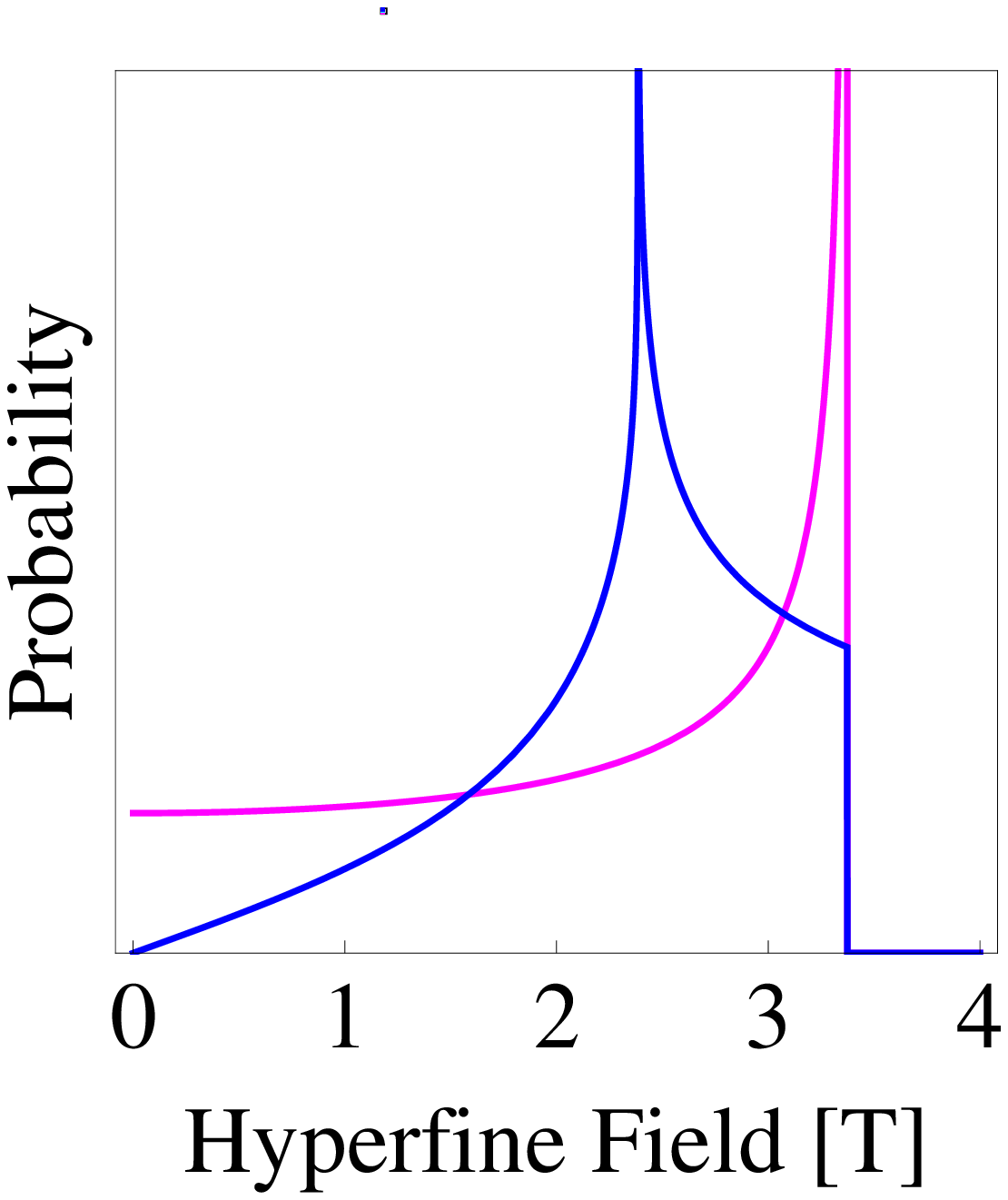}
	\put(-10,95){\textcolor{black}{(a)}}
      \end{overpic}
      \hfill
      \begin{overpic}[height=4cm]{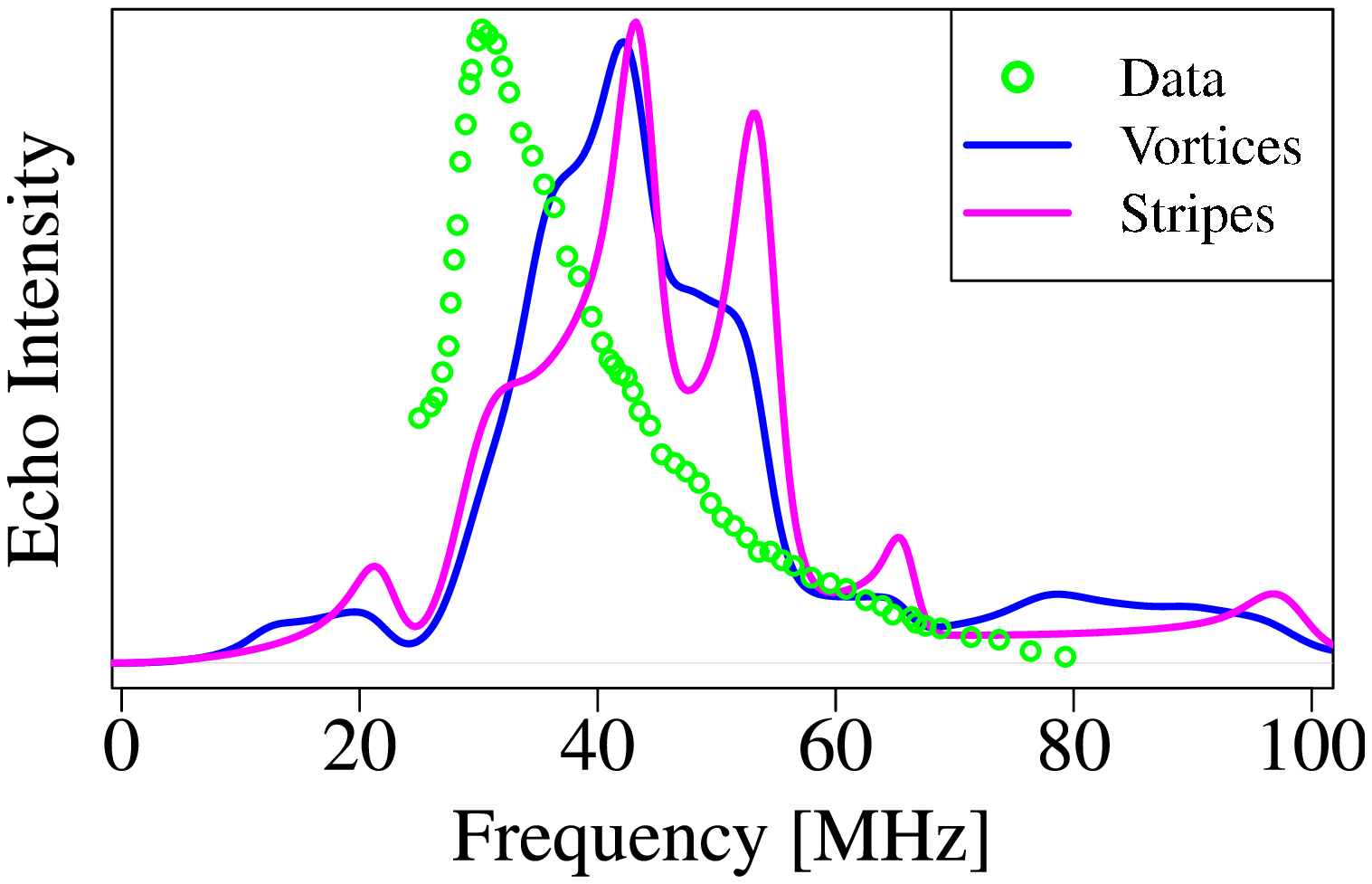}
      \end{overpic}
      \hfill
      \begin{overpic}[height=4cm]{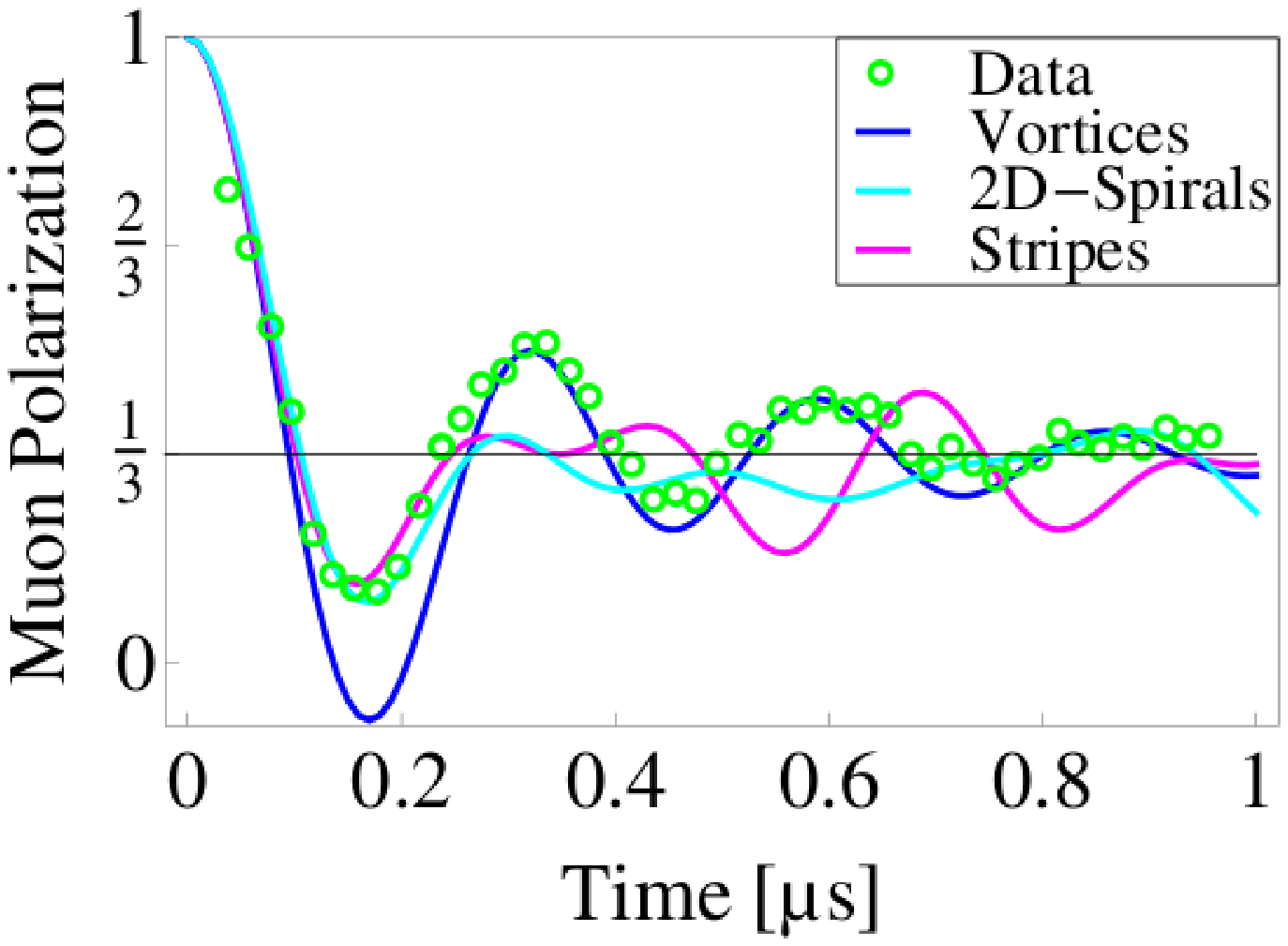}
      \end{overpic}
      \hfill
      \hfill
      \begin{overpic}[height=4cm]{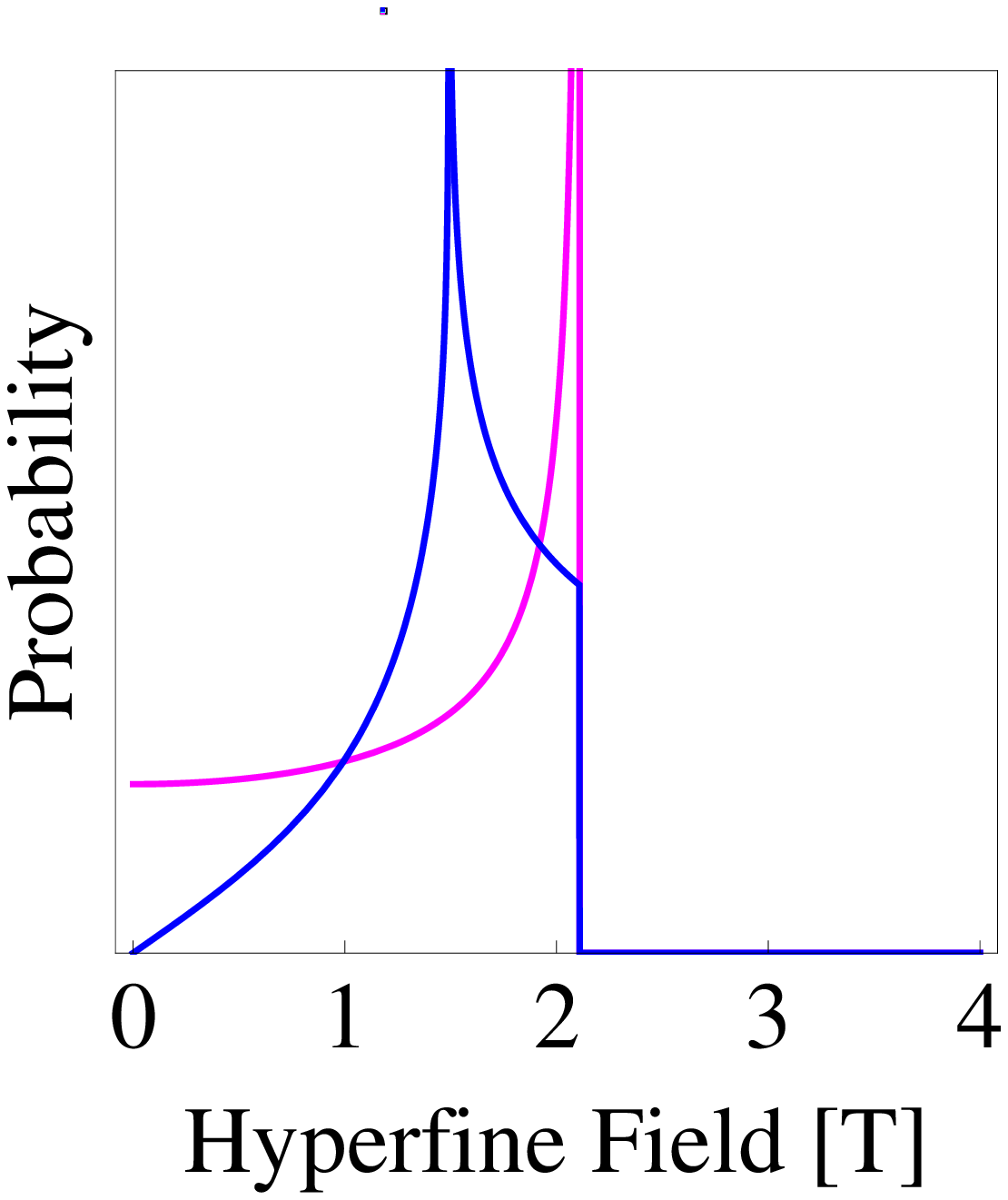}
	\put(-10,95){\textcolor{black}{(b)}}
      \end{overpic}
      \hfill
      \begin{overpic}[height=4cm]{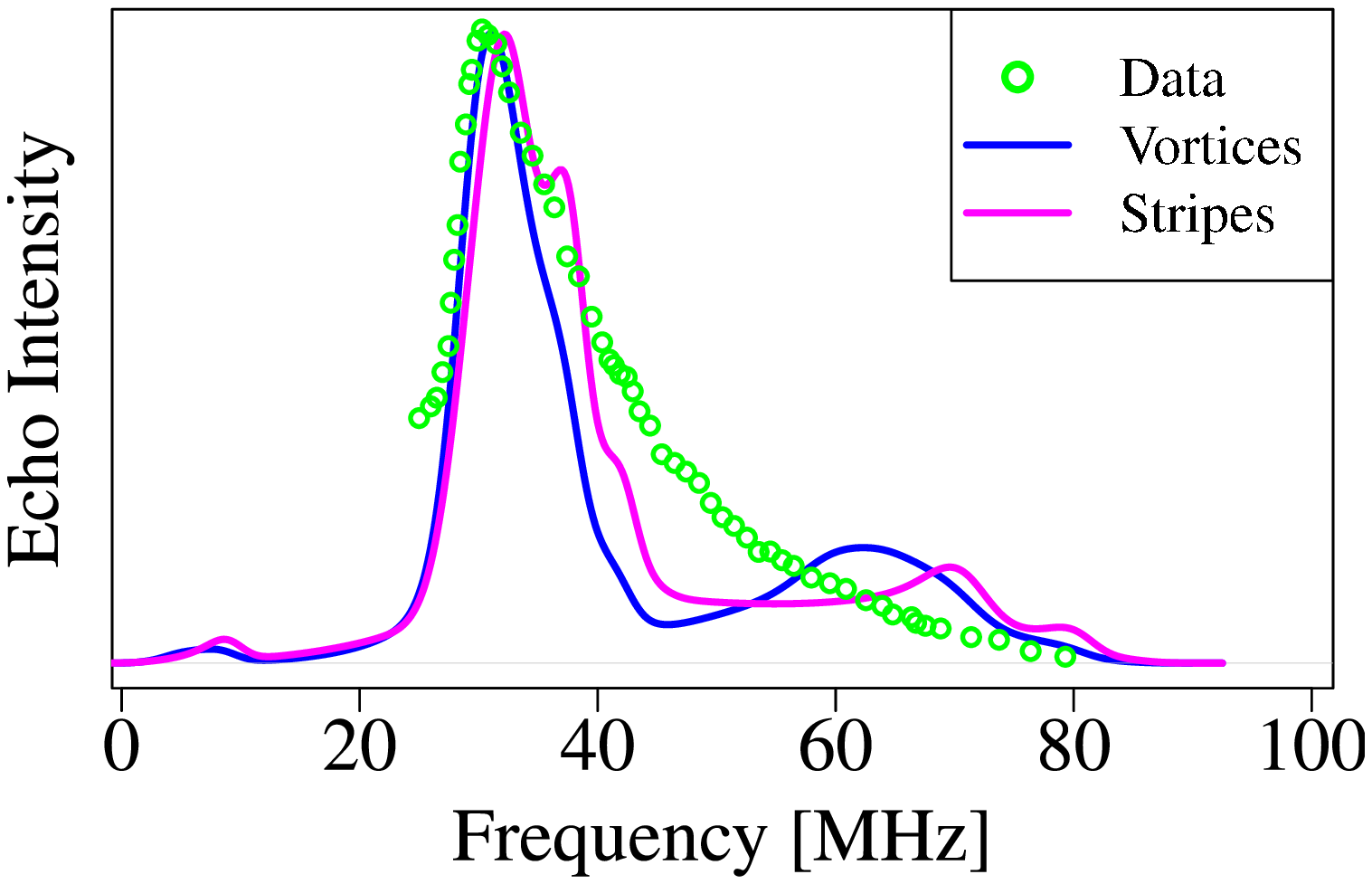}
      \end{overpic}
      \hfill
      \begin{overpic}[height=4cm]{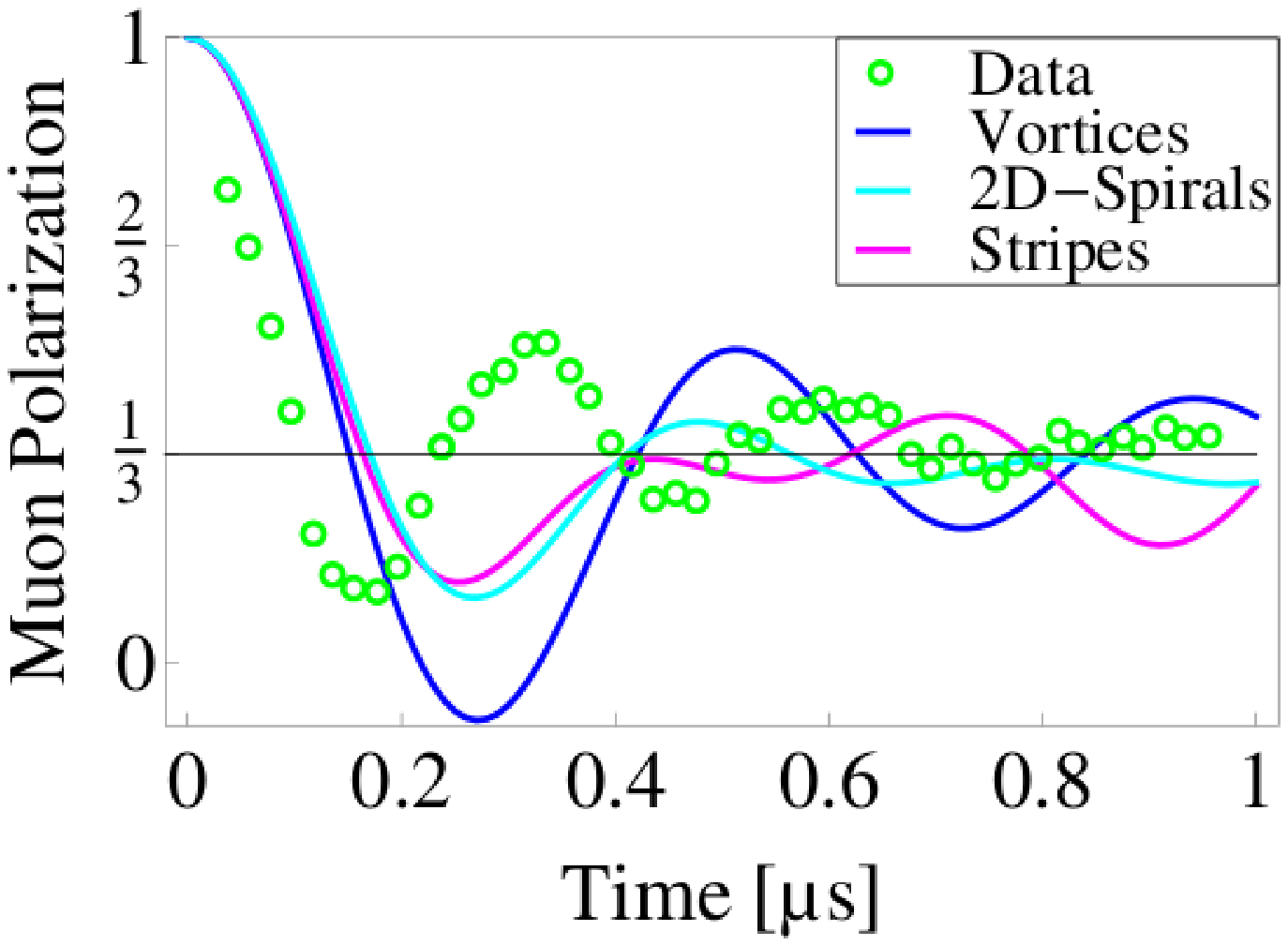}
      \end{overpic}
      \hfill
      \hfill
      \begin{overpic}[height=4cm]{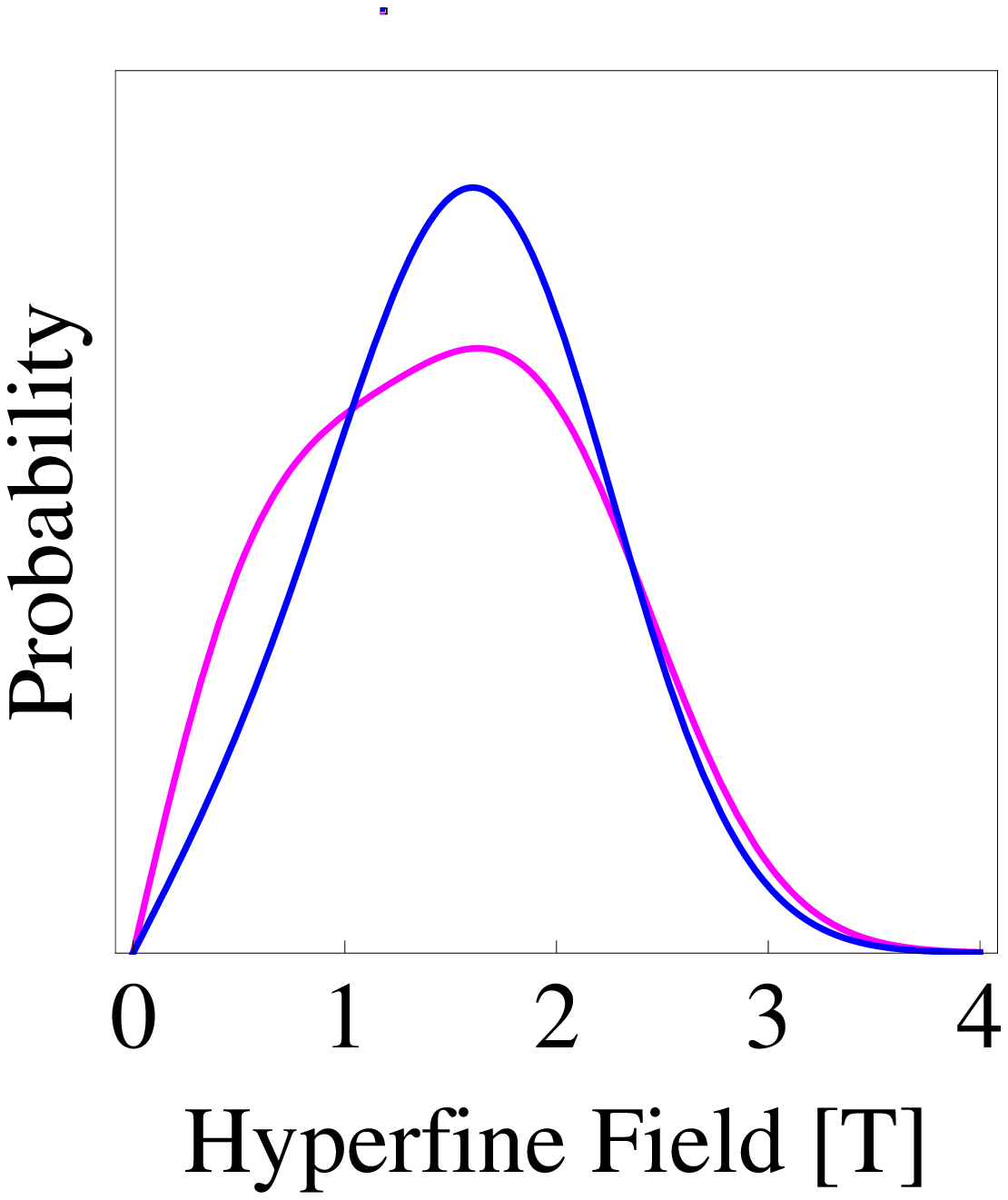}
	\put(-10,95){\textcolor{black}{(c)}}
      \end{overpic}
      \hfill
      \begin{overpic}[height=4cm]{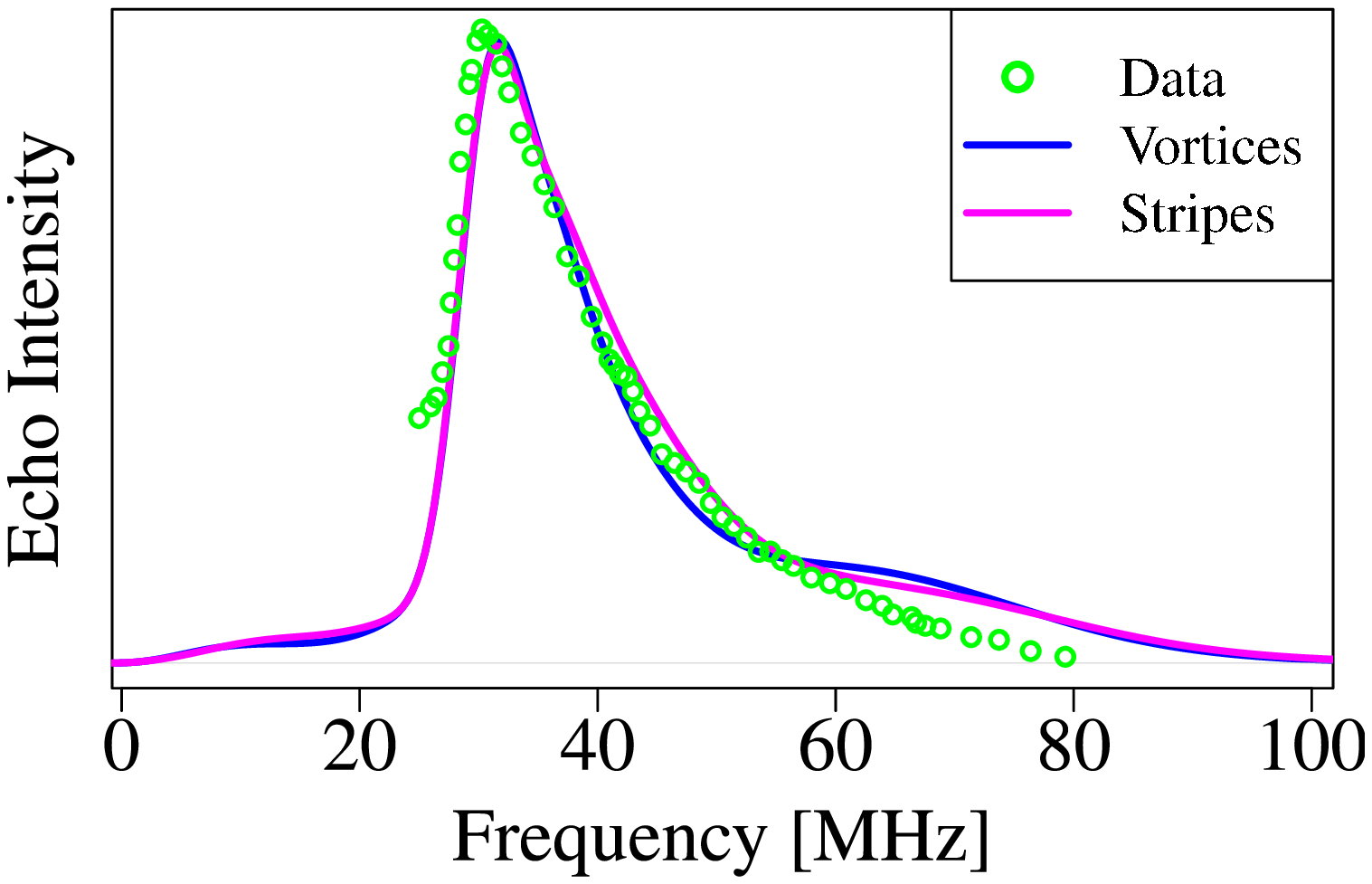}
      \end{overpic}
      \hfill
      \begin{overpic}[height=4cm]{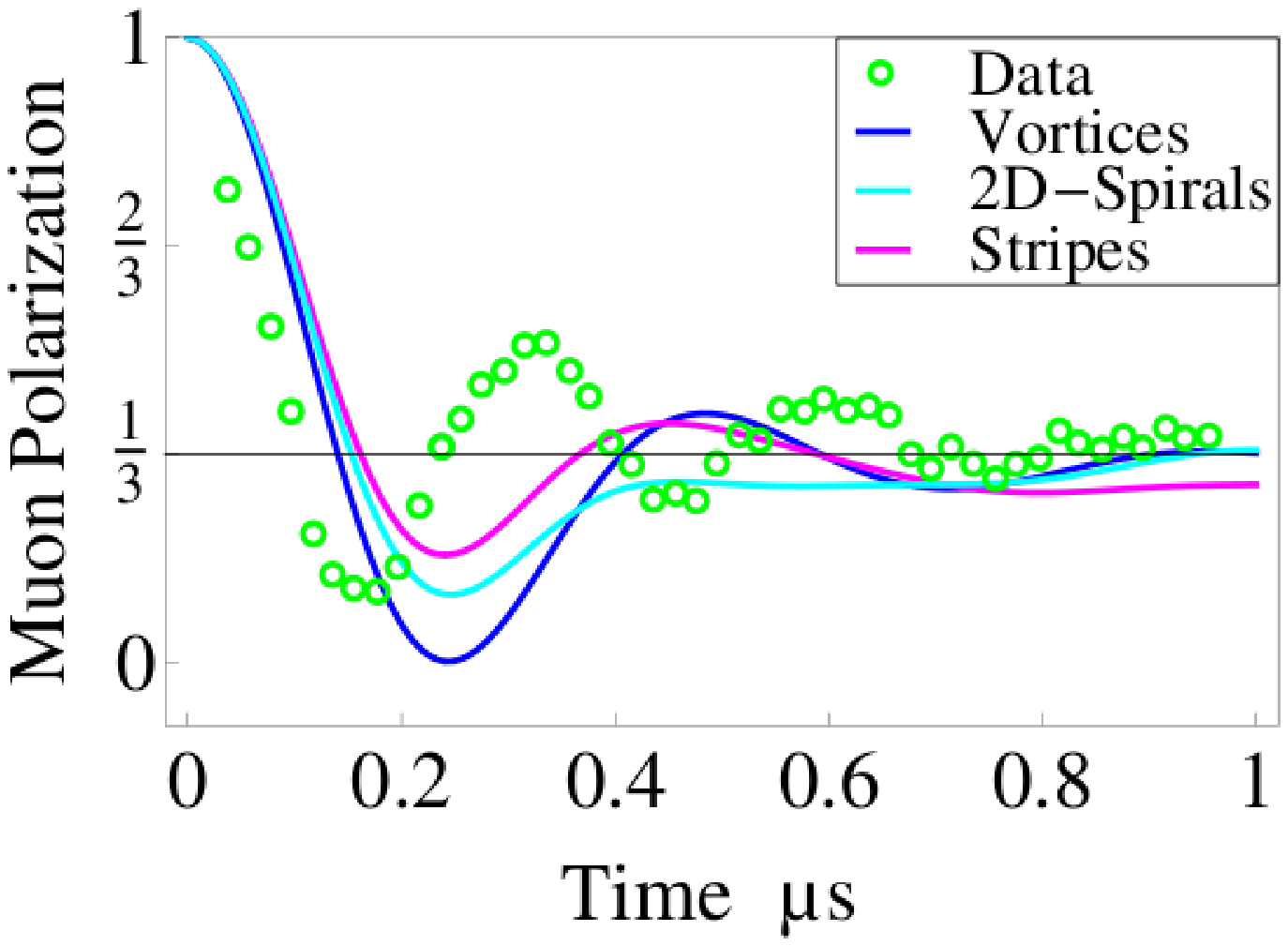}
      \end{overpic}
      \hfill
    \caption{Left column: Distribution of hyperfine fields for an incommensurate spin lattice. Middle column: NQR simulation compared to the experimental data of Hunt \emph{et al.}~\cite{Hunt2001}. Right column: $\mathrm{\mu}$SR simulation  compared to the experimental data of Luke \emph{et al.}~\cite{Luke1997}.
(a)~$S_0=0.32$, $\Delta S=0$, (b)~$S_0=0.20$, $\Delta S=0$, and (c)~$S_0=0.20$, $\Delta S=0.5\,S_0$.
For the left and middle column the curves for the 2D spin spirals coincide with the curves for spin stripes.
}
    \label{NQR}
\end{figure*}

For the spin vortex checkerboard, the resulting distribution of hyperfine fields can than be obtained analytically:
\begin{widetext}
\begin{align}
\rho(H) &=\frac{N_0 H}{H_{m}} \times
\left\{\begin{array}{ll}
\int_{0}^{\frac{H}{H_{m}}} \frac{d \zeta}{\sqrt{(\zeta^2 - 1)\left(\zeta^2 - \frac{H^2}{H_{m}^2}\right) \left( \zeta^2 +1 - \frac{H^2}{H_{m}^2} \right) }} 
 & \textrm{if} \ 0\leq H \leq  \frac{H_m}{\sqrt{2}} \\[0.125cm]
\int_{\sqrt{1 - (H/H_m)^2}}^{1} \frac{d \zeta}{\sqrt{(\zeta^2 - 1)\left(\zeta^2 - \frac{H^2}{H_{m}^2}\right) \left( \zeta^2 +1 - \frac{H^2}{H_{m}^2} \right) }} 	&	\textrm{if} \  \frac{H_m}{\sqrt{2}} < H \leq H_m \\[0.125cm]
0	&	\textrm{elsewhere.}   
\end{array} \right.
\label{FieldDistribution}
\end{align}
\end{widetext}
\noindent Here $N_0$ is the normalization constant, and 
$H_m = S_0[B(2 + \sqrt{2}) - A)$ is the maximum hyperfine field due induced by the spin vortex checkerboard. This distribution has Van Hove singularity at $H = H_m/\sqrt{2}$. 
It is shown in Fig.~\ref{NQR} (a, left column, blue line) and compared to the distribution for the stripe structure (magenta line), which also has a Van Hove singularity but at $H = H_m$.

The distribution of the local hyperfine fields for the two-spiral superstructure is identical to that of the stripe superstructure.

\subsection{${\mu}$SR signal}
\label{Subsec:muSRsimulation}

The time evolution of the muon polarization vector $\textbf{P}_{\mu}$ for polycrystalline samples in the approximation of static local fields and without additional depolarization effects can be calculated as follows
\begin{align}
\displaystyle
P_{\mu}(t) =& \,\frac{1}{3} + \frac{2}{3} \int_{0}^{\infty} p(H)\cos{\left(\gamma_{\mu}H\,t\right)}\,\textrm{d}A
\label{Polarization}
\end{align}
where $p(H)$ is the probability distribution for the absolute value of the local field $H$at the position of the muon and $\gamma_{\mu}=851.61$~[MHz/T] is the muon gyromagnetic ratio. We units of $P_{\mu}$ are to be normalized such that $P_{\mu}(0) = 1$.
We perform the calculations for the muon site identified in Ref.~\cite{Hitti1990} to be $\textbf{r}_{\mu}^{\mathrm{s}}=(0.253, 0, 0.162)$ [in the relative lattice units].

We obtain the distribution $p(H)$ by sampling dipolar fields
\begin{equation}
{\mathbf H} = \sum_{ij} \frac{3(\textbf{S}_{ij}\cdot\textbf{r}_{ij})\textbf{r}_{ij} -r_{ij}^2\textbf{S}_{ij}}{r_{ij}^5},
\label{Dipolfield}
\end{equation}
where $\textbf{r}_{ij}$ is the displacement vector between the muon site and the copper positions on the lattice.
The sampling was done for all possible muon sites in a cluster of $32\times 32\times 16$ spins. 

\section{Results and discussion}
\label{Sec:ComparisonwithExperiments}

\subsection{Joint analysis of NQR and $\mu$SR experiments}

The results of our calculations of NQR lineshapes and $\mu$SR signals are presented in Fig.~\ref{NQR}. The first column of plots contains the distribution of hyperfine fields for the calculation of the NQR lineshape. The second column represents the NQR lineshapes themselves. The third column contains the corresponding $\mu$SR signals.

The first row of panels in Fig.~\ref{NQR} labeled as (a) represents calculations for the modulation amplitude $S_0 = 0.32 \mu_B$ and no spin noise, i.e. $\Delta S = 0$. This choice is most appropriate for fitting the initial behavior of $\mu$SR signals even though significant discrepancies with experiment remain at intermediate and long times. The NQR lineshapes in Fig.~\ref{NQR}(a) for  all modulation patterns considered exhibit apparent disagreement with experiment. The experimental lineshape contains a single peak, which is likely associated with the van Hove singularity in the function converting the hyperfine fields into the NQR frequencies. This van Hove singularity is to be contrasted with the the multiple theoretical peaks appearing in the middle panel of Fig.~\ref{NQR}(a) and originating from the van Hove singularities in the distribution of the hyperfine fields themselves [see the left panel of Fig.~\ref{NQR}(a)]. 

The second row of panels in Fig.~\ref{NQR} labeled as (b) represents our best attempt to fit the NQR lineshape by reducing the amplitude of spin modulations, still without introducing spin noise. The optimal modulation amplitude was $S_0 = 0.2 \mu_B$. For this value, the principal peaks originating from the van Hove singularities shown in the left panel of Fig.~\ref{NQR}(b) coincide with the peak of of the function converting the hyperfine fields into the NQR frequencies. The experimental NQR line is still noticeably broader then either of the theoretical predictions. The corresponding theoretical $\mu$SR signals are noticeably slower than the experimental one. 

Finally, the third row of panels in Fig.~\ref{NQR} labeled as (c) represents our best attempt to further improve the agreement with the experimental NQR lineshape by adding spin noise: $S_0 = 0.2 \mu_B$, $\Delta S = 0.5 S_0$. In this case, all three patterns considered lead to a good agreement with the experimental NQR shape, but a significant discrepancy with the experimental $\mu$SR results remains.

The overall outcome of our calculations is that neither of the patterns considered can accurately account for both the NQR and the $\mu$SR experiments with the same values of $S_0$ and $\Delta S$. Either of the patterns can well reproduce the NQR lineshape with a smaller modulation amplitude and in the presence of a strong degree of spin noise. The $\mu$SR experiments are best fitted with a larger modulation amplitude and in the absence of the spin noise. Still, even the best fits of the $\mu$SR experiments exhibit significant disagreements with the experiments. 

Our experience suggests that the origin of the above persistent disagreement in the shape of the $\mu$SR signal  is beyond the specific choice of the spin modulation pattern. We expect a significant disagreement for any pattern of the static local spin configurations. It originates from the fact that the three-dimensional arrangement of static spin polarizations unavoidably leads a three-dimensional distribution of magnetic dipole fields experienced by muons. For a three-dimensional distribution, the probability of the zero absolute value of magnetic field is normally equal to zero.  This probability is proportional to the Fourier transform of $P_{\mu}(t) -1/3$ at zero frequency. Therefore, zero probability for the zero absolute value of the local magnetic field implies that $\int_0^{\infty} [P_{\mu}(t) -1/3] dt = 0$. The experimental $\mu$SR signal used for comparison with theory in Fig.~\ref{NQR} apparently does not satisfy the above condition.
It is not clear to us whether the above discrepancy has something to do with the inadequacy of the theoretical assumption of static local fields, or with experimental uncertainty in terms of identifying the 1/3-level of the $\mu$SR signal. 

Irrespectively of the above disagreement, it was argued by Kojima et al.\cite{Kojima2000}, that large amplitude of $\mu$SR oscillations indicate that the spin structure is one-dimensionally modulated. This conclusion was based on the the comparison between the results for one-dimensional stripes and two-dimensional grid. However, our calculations indicate that, in the case of the two-dimensional spin-vortex checkerboard, the resulting oscillations are even larger than for the stripe pattern. This indicates that large amplitude of $\mu$SR oscillations as such cannot discriminate between stripes and spin vortices and hence between 1D and 2D spin patterns.

Finally, we have attempted to explore the possibility that the muon site is different from the one identified by Hitti et al.\cite{Hitti1990}. Trying the muon positions in the $ac$-plane 1\AA \ away from the apical oxygen, we were able to improve the $\mu$SR fit  for the stripes and 2D spirals with muon position at $\textbf{r}_{\mu}^{\mathrm{s}}=(0.110, 0.0, 0.114)$   and with $S_0 = 0.15 \mu_B$, $\Delta S = 0$. In this case, the NQR spectrum has a single main peak, which is much narrower than the one observed experimentally, but with the maximum at the experimentally observed frequency. No comparable improvement of the $\mu$SR fit was found for the spin vortex checkerboard. 

\

\section{Conclusions}
\label{Sec:Conclusion}

The outcome of the analysis presented in this paper indicates that spin vortex lattice leads to the singularity in the distribution of local magnetic fields, which is quite similar to the one produced by the stripe pattern. As a result no clear qualitative differences emerge between the NQR/$\mu$SR predictions based on stripes or spin vortices, while, in the both cases noticeable quantitative discrepancies with the experimental results remain. Similar discrepancies also appear for  the 2D superposition of spin spirals. It is only clear that the observation of the large amplitude of $\mu$SR oscillations does not amount to a qualitative argument to support stripes against the spin vortex checkerboard.

\end{document}